# Where The Light Gets In: Analyzing Web Censorship Mechanisms in India

Tarun Kumar Yadav, Akshat Sinha, Devashish Gosain, Piyush Sharma, Sambuddho Chakravarty
{tarun14110,akshat14132,devashishg,piyushs,sambuddho}@iiitd.ac.in

Indraprastha Insitute of Information Technology Delhi, New Delhi, India

## ABSTRACT

This paper presents a detailed study of the Internet censorship in India. We consolidated a list of potentially blocked websites from various public sources to assess censorship mechanisms used by nine major ISPs. To begin with, we demonstrate that existing censorship detection tools like OONI are grossly inaccurate. We thus developed various techniques and heuristics to correctly assess censorship and study the underlying mechanism involved in these ISPs. At every step we corroborated our finding manually to test the efficacy of our approach, an exercise largely ignored by several others. We fortify our findings by adjudging the *coverage* and *consistency* of censorship infrastructure, broadly in terms of average number of network paths and requested domains the infrastructure surveils. Our results indicate a clear disparity among the ISPs, on how they install censorship infrastructure. For instance, in Idea network we observed the censorious middleboxes on over 90% of our tested intra-AS paths whereas for Vodafone, it is as low as 2.5%.

We conclude our research by devising our own novel anti-censorship strategies, that does not depend on third party tools (like proxies, Tor and VPNs *etc.*). *We managed to anti-censor all blocked websites in all ISPs under test.*

## 1. INTRODUCTION

Free and open communication over the Internet, and its censorship, is a widely debated topic. It is not surprising that an overwhelming majority of prior studies on censorship activities and their mechanism, primarily center around overtly censorious nations like China [27, 40, 34, 52] and Iran [24]. Most of these studies involve reporting censorship activities, with some categorically focusing on the in-depth description of the actual censorship techniques and mechanism that are employed by such nations; *viz.*, describing the network location of the censorship infrastructure, what triggers them and how are clients notified of such filtering (*i.e.* if at all).

Through our studies over the past few years, we discovered that even democratic nations like India, have slowly, and rather covertly, evolved as an infrastructure for large-scale Internet censorship, involving several privately and federally operated ISPs. India's Internet censorship policies have remained arbitrary (at best ambivalent)[1]. Over time several networks have upped their barriers against users accessing sites, which the administration "believes" to be "unfit for consumption", resulting in enough citizens facing web censorship.

Rather than analyzing the present censorship infrastructure and policies, previous work [22] emphasized on hypothetical scenarios of potential (future) large scale censorship (or surveillance) by the state. A mere preliminary report was also presented highlighting the inconsistent web censorship policies amongst ASes.

We thus formally approached the authorities, filing a *Right to Information* [15] request (RTI), inquiring about the policies and mechanism the government uses to block content. In response, the authorities shared that while the censorship policies are confidential, the onus of implementing them lied with the individual ASes who could employ any mechanism they chose.

Unambiguous answer from authorities motivates us to conduct our own detailed analysis of the different censorship mechanisms the major network operators of the country employ. We began our research by compiling a corpus of about 1200 potentially blocked sites (PBWs), curated from various Internet sites (*e.g.* Herdict [6], Citizen Labs [10]). Thereafter we obtained network connections for nine popular ISPs.

For assessing censorship, we ran the popular censorship assessment tools like OONI [18] on clients hosted in these networks. OONI runs two sets of tests, one at the client and other at their remote control site (assumed to be unfiltered). A mismatch between the results signals potential censorship. However, our initial tests yielded considerably high false positives and negatives when tested through different ISPs. For instance, in Airtel, we obtained a false positive rate of $\approx 80\%$ and a false negative rate of $\approx 11.6\%$.

We thus decided to devise our own analysis techniques. We began by observing the ensuing network connection traffic between our client and the censored

---

[1]For *e.g.* in August 2015, the government issued orders to block 857 websites, but later backtracked under public outcry [13]

site. In one particular ISP network, we observed that whenever the client connected to the censored site, it received a valid HTTP response bearing a statutory censorship notification with appropriate sequence number and bits (*e.g.* FIN, RST) in the TCP headers that enforce the client to disconnect with the server. Eventually, the actual response from the censored site also arrives, but by then the connection is already terminated, and the packet is discarded.

All such protocol exchanges hinted toward the presence of malicious network elements (we collectively call *middleboxes*) that snoop (or intercept) users' traffic and upon observing requests to filtered websites, injects the aforementioned crafted packets to censor traffic.

To identify the network location of such censorship infrastructures, we devise a technique which we collectively call *Iterative Network Tracing*, that works on the principle used by `traceroute`. It is quite similar to those proposed earlier by Xu *et al.* [52] and involves sending web requests to censored sites but with increasing IP header TTL values, so that en route the messages encounter middleboxes, that are triggered upon the arrival of request to the censored sites.

Using our approach, Iterative Network Tracing, and and various heuristics which we developed after observing peculiarities of censorship techniques, we conducted an investigative study of various censorship mechanism, employed by major ISPs in the country. Our research engages long-term data collections to answer the following questions:

- What sequence of protocol messages triggers censorship?

- Exactly what techniques are employed by ISP networks to filter users' requests to censored sites?

- Approximately what fraction of network paths are impacted by these censors?

- Is censorship uniform and consistent across the various ISPs? More specifically –

  - Do various ISPs block the same set of sites?
  - Do various censorship devices of an ISP (*aka* the *middleboxes*) block the same set of sites?

- How hard or easy is it to bypass such censorship mechanism?

Unlike several previous efforts, that directly draw conclusions based on the results generated by their respective tools and techniques [30, 31, 44] ours, at every possible step, involves corroborating the results via manually connecting to the sites and inspecting the results.

The key contribution of our research efforts, spanning over 18 months, involves detailed answers to all the aforementioned questions. Our findings show that four of these ISPs *viz.* Airtel, Vodafone, Idea and Reliance Jio (potentially carrying a large fraction of network traffic [22]), employ stateful inspection of HTTP requests alone to censors access. For some ISPs, like Idea Cellular, we detected the presence of censorship infrastructure in over a very large fraction (>90%) of the intra-AS network paths.

Others, *viz.* BSNL and MTNL, prime government operators, poison DNS responses for censored sites. In our experiments, we identified about 600 censorious DNS resolvers spread across these two ISPs.

Traffic of non-censorious ISPs transiting the censorious ones often gets inadvertently filtered. Collectively, this is known as *collateral damage* [39]. We observed such phenomenon for various non-censorious ISPs in India. For example, censorship in Vodafone network causes collateral damage to NKN, an otherwise non-censorious educational network.

Through our detailed explorations, we discovered network middleboxes that either intercept traffic (like transproxies) or merely snoop on users traffic and sends back specially crafted messages disconnecting the client–server connection. While a vast majority of previous efforts, like [35, 42, 49] report the latter, to the best of our knowledge we are the first ones to discover the presence of intercepting middle box in any ISP.

Finally, while relatively powerful censorious nations have evolved mechanism to counter censorship circumvention mechanism [17], we demonstrate simple, yet effective, mechanisms that can be used to bypass censorship without relying on popular anti-censorship tools like proxies or VPNs; making them harder to bypass approaches proposed by others. Our approach relies on identifying the packets generated from the middleboxes and filtering them at the client, or sending crafted requests that go undetected by the middleboxes, but are correctly recognized by the server. This is harder for ISPs to identify that work by restricting access to anti-censorship infrastructure. Moreover, efforts to retrofit the solution into existing censorship middleboxes may incur high costs on the part of the middlebox manufacturers and the ISPs, without factoring in downtimes and potential failures.

## 2. BACKGROUND AND RELATED WORK

According to Open Net Initiative report, India is among the list of countries that restricts the Internet content and ranks India as "Partly Free" [9]. Internet censorship in India can be traced back to the year 1999, where website of the popular Pakistani daily newspaper 'Dawn' was blocked from access within India, immediately after the Kargil War [14]. Since then, there are numerous instances of Internet censorship recorded [9] by the orders of the government to an extent of Internet Shutdowns. In the year 2015, there were at least 22 instances of Internet shutdowns in different parts of the country [50]. And later in the same year, Internet service providers (ISPs) have been asked by the Government of India to block 857 websites, on the basis of restricting access to pornographic content [3].

Very recently transparency reports published by Facebook [4] and Google [5] also confirm that censorship in India is on the rise. It indicates there were a total of 21 instances of complete Internet shutdowns and 1,228 instances of content removal by Facebook because a majority of content restricted was alleged to violate local laws relating to defamation of religion and hate speech.

Thus, we conducted a detailed study of web censorship trends pertaining to Indian ISPs. Specifically, we aim to explore the censorship mechanism and its associated infrastructure deployed in the country.

We begin by discussing important studies in the area of Internet Censorship, primarily reporting the *type* and *mechanism* of censorship. Zittrain [55] in his seminal analysis of censorship observed IP, keyword and DNS filtering in China. Later many studies focused on censorship specific to particular countries for *eg.,* China [51, 27], Pakistan [42], Italy [20], Greece [48] Iran [24], Egypt and Libia [28] etc. Verkamp *et al.,* [47] extended this work by deploying clients in 11 countries to identify their network censorship activities encompassing IP and URL filtering, keyword filtering and DNS based censorship etc. Gill *et al.,* [33] rather than deploying clients, used data gathered by the OpenNet Initiative to detect censorship in 77 countries.

Dalek *et al.,* [29] used data from Shodan [16] to identify URL filtering products deployed across many countries including Qatar, Yemen, Saudi Arabia and India. For large scale detection of censorship across multiple countries, there are several projects which provide tools to determine censorship policy: HerdictWeb [54], CensMon [45], and Encore [25], OONI [19] and Augur [43]. However, a significant portion of censorship literature focuses only on the Republic of China — Great Firewall of China (GFW) [55, 27, 53, 52, 41, 30, 31, 49, 38]. Winter *et al.,* [17] studied how DPI-enabled routers detect Tor bridges based on specific TLS cipher suits. Others such as [39] reported that China is heavily contributing towards collateral damage by DNS filtering. Khattak *et al.,* [37] observed that GFW operates similarly to NIDS and found exploitable flaws in state management of GFW. Later authors in [49], reported that GFW has evolved over a period of time and previous solutions [37] to bypass it, have failed. They proposed a novel tool INTANG, to anti-censor GFW using carefully crafted packets, without relying on third-party software bundles like Tor and VPNs.

In the year 2017, we [22] explored that Indian ISPs have incoherent censorship policies and they implement their own content filters resulting in dramatic differences in the censorship experienced by customers. Also, we studied the hypothetical scenario — assuming in future, the government of India plans to implement strict censorship what would be the probable 'key points' to place the filters, for different censorship mechanisms *viz.,* IP filtering, Prefix Hijack, DNS filtering etc.

In this research, we rather carried out a comprehensive study on the '*present*' Internet censorship implementation in India, which was missing in our previous analysis.

## 3. DATA COLLECTION AND APPROACH

For our research, we curated a list of potentially blocked websites (which we conveniently refer to as PBW) from different sources which include Citizen Labs [10], Herdict [6] and various past government and court orders of the country [11]. The list includes a total of 1200 websites which we consider to be sensitive (and thus potentially censored). They span across 7 major categories *viz.,* escort services, pornography, music, torrent sites, politics, tools and social networks.

We commenced our research by using the already available tool OONI [18]. A client which intends to detect possible instances of censorship (at different layers of network stack) installs OONI probe. This fully automatic tool, after running for finite time, reports the blocked websites and possible censorship mechanisms behind them. After running OONI from five different vantage points we observed that it results in high false positives and negatives. Thus, we created our own scripts to detect Internet censorship in India.

### 3.1 The OONI tool

*Open Observatory of Network Interference* (OONI) [19], is an open source tool under the Tor project and is designed to detect censorship.

We ran OONI on five popular ISP networks, using the PBW, and recorded the results. To corroborate our findings we also manually checked the sites that were reported by OONI as being censored. To our surprise, we observed very few true positives. An exceedingly large number of sites which were reported as being censored was however easily accessible.

Table 1 summarizes our findings. The rows represent the ISPs, columns correspond to the type of censorship reported by OONI, and each entry is a 2-tuple (P,R) representing the precision and recall.

To explain our results better we use the example of data gathered for Airtel (a major ISP of the country[2]). The OONI tool reported that about 78 sites ($B_O$) were being blocked by the Airtel. Upon manual inspection we observed this number to be much higher, *i.e.* 133 ($B_M$). Only 15 websites ($B_O \cap B_M$) that were actually being censored were also corrected detected by OONI. This provides us with a precision of 0.19 ($|B_O \cap B_M|/|B_O|$) and a recall of 0.11 ($|B_O \cap B_M|/|B_M|$). Similar results were observed for other ISPs as well.

---

[2]In terms of network paths it intercepts [22]

| Popular | Censorship Type | | | |
|---------|---|---|---|---|
| ISPs | Total | DNS | TCP | HTTP |
| *MTNL* | 0.57, 0.42 | 0.44, 0.10 | 0, 0 | 0.60, 0.64 |
| *Airtel* | 0.19, 0.11 | 0, 0 | 0, 0 | 0.19, 0.11 |
| *Idea* | 0.57, 0.62 | 0, 0 | 0, 0 | 0.57, 0.62 |
| *Vodafone* | 0.69, 0.82 | 0, 0 | 0, 0 | 0.70, 0.78 |
| *Jio* | 0.34, 0.15 | 0, 0 | 0, 0 | 0.36, 0.14 |

Table 1: Accuracy of OONI: Precision and Recall values, measured in various ISPs.

Such low values of precision and recall can be attributed to the fact that OONI tool uses rudimentary approaches to detect potential censoring activities. For instance, while detecting DNS filtering, it compares the IP address of a given host name returned by Google DNS resolver (which they assume to not be tampered) with the IP address mapped to that website by the client's ISP. If the two IP addresses of the same website are different they assume it to be censorship. But, in many cases differences in URL resolution is likely an artifact of network hosting architectures (*e.g.* CDNs).

Also, while detecting HTTP filtering, OONI sends an HTTP request to a given website over the network where the client (running the OONI probe) is hosted. Following it, the same request is sent from the control server (of OONI). HTTP responses obtained from these requests are compared (based on a threshold) and the website is assumed to be filtered if the responses differ. However, while doing our experiments, in many cases we observed that in spite of having the difference in the results, websites are not blocked (explained in detail in section 6).

Thus, for our research, we abandoned OONI and created our own *semi-automatic* scripts to record the censorship instances by various ISPs across India. For instance, similar to OONI, we tried sending `GET` requests to PBWs from the client in test ISPs and through Tor. If the difference in responses is less than the threshold we considered them *non-censored*, otherwise we manually inspect the responses further, unlike OONI, which directly flags them as censorious. For instance, in Airtel network, when we selected the threshold as 0.3, we observed that difference between aforementioned HTTP responses for 390 websites, were more than the threshold. On manually verifying the contents we observed that 40% of 390 websites were still non-censored[3]. We repeated the same experiments for all test ISPs under consideration and found that $30 - 40\%$ of the websites which would have been flagged as censorious by OONI are actually non-censorious.

---

[3]This difference between the responses is further explained in section 6.

We now present our approach for determining the type of censorship (DNS, TCP/IP and HTTP blocking) and mechanisms behind them.

## 3.2 DNS Blocking

**I. Background:**. For ordinary netizens, DNS resolution is the primary step for accessing any website. URL entered by such netizens is first resolved to its associated correct IP address. Thus, invariably censors exploit this step, and often return an incorrect IP address, resulting in website's unavailability.

DNS based blocking can be achieved by (1) *DNS poisoning* [46]— whereby a local resolver is corrupted and replies with the incorrect IP for the specific DNS queries (2) *DNS injection* [23] — where some middlebox between the client and local resolver intercepts the DNS query and deliberately responds with an incorrect IP address.

**II. Identifying filtered domains:**. In order to identify DNS filtering by ISPs, we selected PBWs that could otherwise be successfully resolved through Tor circuits (ending with exit nodes in non-censorious nations). Thereafter, we attempted to resolve these PBWs through ISPs that we chose to test for potential censorship activities.

A URL may resolve to different IP addresses, depending on the network from which resolution is attempted (due to reasons such as CDN based hosting). We identified URLs whose resolution resulted in an overlapping set of IPs when the resolution was attempted both by the client (hosted in the ISP under consideration), as well as via the Tor circuit. The sites corresponding to such overlapping sets of IPs were considered uncensored.

In order to ascertain DNS filtering on the remaining list, we first tested our intuition *i.e.,* an ISP, may resolve multiple blocked websites to some unique IP addresses. Thus, we performed frequency analysis on the observed IP addresses *i.e,* if more than one website is resolved to same IP address[4], we checked it for DNS filtering. Invariably we found static IP address of the same ISP appearing multiple times. We also observed several bogon IP [1] addresses.

Thus, for the remaining URLs, that resolved to non-overlapping sets of IP addresses, we applied the following heuristics to decide if they corresponded to responses that is seemingly manipulated by the censor.

1. *Resolved IPs belong to the same AS that hosts the client:* None of the PBWs were hosted in the clients' AS. Thus if any of the resolved IPs belong to the clients' AS (the one under test), the AS is considered censorious and the corresponding URL is marked censored.

---

[4]We initially scanned the list and removed those cases for which multiple websites were actually hosted on the same IP address.

2. *Resolved IPs are Bogons*: If any of the resolved IPs is a bogon [1], we consider the AS to be censorious and the URL as being censored.

In order to ensure that *only* the aforementioned heuristics are applied by our tested ISPs, to the remaining IPs we sent the HTTP request through Tor circuit. We manually confirmed that content was received from all the IP addresses, indicating that all the remaining IPs were correctly mapped to the websites.

**III. Identifying DNS filtering mechanism:**. After identifying the set of websites which are censored due to DNS filtering, we intended to identify the mechanism behind the blocking. To that end, we began by identifying all open DNS resolvers of the ISP under consideration. To do this we send DNS queries requesting resolution for otherwise uncensored sites (*e.g.* our own institution's website) whose correct IP address is known beforehand, to the entire IPv4 address space of the said ISP. DNS resolvers, even if censorious but otherwise configured correctly, are expected to respond to such queries with legitimate IP addresses.

In order to identify only the *censorious resolvers*, we sent 1200 DNS queries, corresponding to the individual PBWs, to each of the DNS resolvers (in each of the individual censorious ISP). Resolvers which responded with manipulated IP address (even for one DNS query), are considered to be censorious.

In order to determine **how and where censorship happens**, *i.e.* DNS injection by middleboxes or manipulated responses by the poisoned resolvers involves a variant of our Iterative Network Tracer (as shown in figure 1). We began by identifying the router-level path between them beforehand using `traceroute`. Thereafter, the client sends DNS requests (corresponding to PBWs) to *only* censorious DNS resolver by iteratively increasing IP TTL values. Identifying censorship mechanism would involve checking if the responses (between the client and a PBW) arrives from any network hop other than the last one. Responses from network hops other than the last ones are likely due to middleboxes, else they are due to poisoned resolvers.

In all our tests we received manipulated IP addresses from the last hop only, indicating the presence of **DNS poisoning**.

### 3.3 TCP/IP Packet Filtering

Network protocol header based filtering is a rather ill defined, albeit very commonly assumed, network censorship technique. It is frequently believed by netizens that ISPs filter traffic based on IP addresses and port numbers. Not surprisingly several past research efforts focus on detecting censorship where they claim that ISPs filter traffic based on IP addresses. To that end, they primarily rely on packet drops corresponding to TCP connection attempts [43] and claim them to be due to IP level censorship.

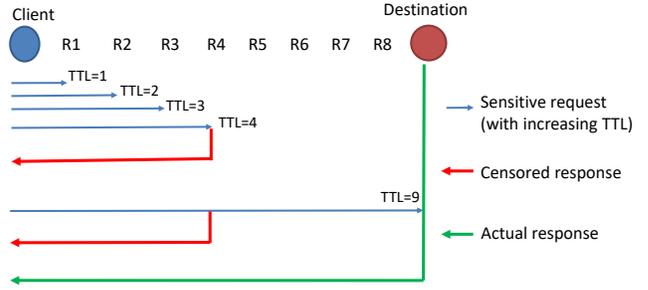

**Figure 1: Iterative Network Tracer: client sends a crafted query (DNS query/HTTP `GET` request) containing a blocked domain with increasing TTL. Censored response is observed from the malicious network element.**

In general, IP address based censorship may be associated with various kinds of systemic failures such as network congestion and outages, router disconnections and delays in route re-computations and convergence. More, it may be hard to distinguish IP address based censorship from local firewalling by the site operator, as that may also result in packet drops. Finally, unlike HTTP censorship, which often involves users receiving censorship notification packets, IP address filtering drops reveal no information to the client, and become hard to distinguish it from the other reasons mentioned. Such scenarios are very difficult to validate (an important, and often ignored aspect of prior research [43]).

Nevertheless, we used a rather crude approach to detect filter based on network and transport protocol headers. We attempted a TCP $3-way$ handshake using a Tor circuit, terminating in a non-censorious country, to the PBWs. For those websites where connection succeeded via Tor, we again attempted five subsequent TCP $3-way$ handshakes (from the ISP under test) with a delay of approximately two seconds between each of them. If it failed in *all* attempts, it implies TCP/IP filtering. However, in none of the ISPs, we ever obtained this form of censorship.

### 3.4 HTTP Filtering

**I. Background:**. HTTP filtering aims at hampering the communication between client and server by observing the content of HTTP packets. The censor can achieve this type of filtering by deploying middleboxes in the network (placed between the client and the blocked domain).

**II. Detecting HTTP filtering:**. In our experiments, we tested all our ISPs for potential censorship using our curated list of 1200 PBWs. We began by creating Tor circuits terminating in non-censorious countries. Through these, we tried accessing all the PBWs. The retrieved contents were compared against the contents obtained by connecting directly to same respective PBWs, directly from our clients hosted in the individual ISPs.

We calculated the HTTP diff [5] between the aforementioned responses and if the difference is less than the threshold (0.3 in our case), we consider it to be non-blocked otherwise else we manually inspected all the responses to remove false positives/negatives.

**III. Which HTTP messages trigger censorship?.** We initiated our study of determining what triggers censorship by observing the protocol messages between client and PBWs. For *e.g.*, for a client hosted in Airtel, we observed that as soon after it sends an HTTP GET request to a PBW (following a regular TCP 3-way handshake), a HTTP 200 OK response packet arrives, whose source IP address is that of the PBW. It had the TCP FIN bit set and payload carried the censorship notification. TCP FIN bit forced the client's browser to initiate TCP 4-way connection termination with the PBW. Eventually, the packet from the PBW *also* arrived. We were thus unsure as to what triggered censorship – requests from the client to the PBW or their responses?

In the past researchers have reported evidence of middleboxes in China that inspect both *request* from the client and *response* from the server for censoring content [26, 27]. Thus, we also required heuristics to determine whether censorship was triggered by requests or by responses.

In order to distinguish between the two possibilities, we adopted the following approach. Initially, the client runs `traceroute` to obtain the number of hops ($n$) to the actual website. Thereafter, we establish a TCP 3-way connection with the website and send two consecutive HTTP GET requests for the blocked website. The IP header of the first request has a TTL value of $n-1$ and is not expected to reach the site, and thus no responses from the site are expected. Whereas, the second is sent to the site bearing a TTL value of $n$ (and is expected to be handled like a regular request).

Depending upon what the middleboxes en route inspect, there could be three possibilities:

- Possibility 1 (Middlebox inspects *only* the request): Both the above requests would traverse the middlebox and would respond back with a censorship notification-cum-disconnection message.

- Possibility 2 (Middlebox inspects *only* the response): The middlebox would be triggered when it inspects the response messages, which happens only when the request actually reaches the site and elicits it (*i.e.* only corresponding to the second request).

- Possibility 3 (Middlebox inspects both *request and response*): The middlebox sends censorship notification-cum-disconnection message for both the request.

In our measurements for all ISPs, we observed censorship notification-cum-disconnection packet for both the requests (*i.e.* for TTL=$n-1$ and TTL=$n$). This directly rules out the possibility 2, *i.e* middlebox getting triggered *only* through responses.

The only remaining possibility is possibility 3, wherein the middlebox inspects both requests and responses. In order to distinguish possibilities 1 from 3, we crafted our own HTTP GET request such that PBW interprets it correctly but *not* the middlebox. For *e.g.*, in Airtel network, merely manipulating the case of the HTTP header field Host and changing it to HOST was sufficient for the request to go undetected by the middlebox (while be correctly interpreted by the PBW). We show in section 5 that for all ISPs, we managed to bypass the censorship by only modifying the HTTP header fields of the GET request. *This confirms that middlebox is only inspecting the request (possibility 1) and not the response, as otherwise, we would still be receiving censorship notification when the responses carrying potentially censored content would be intercepted (by the said middlebox)*.

**IV. How GET request triggers the middlebox?:**. Since middleboxes inspect the GET request for potential censorship, we intend to confirm exactly how the middleboxes get triggered. By default a regular GET request bares only the domain name along with the requested page. We first ran `traceroute` to obtain the number of hops to the server. Then, we crafted a GET request whose IP TTL was set to the value of penultimate hop, such that it passes the middlebox but never reaches the server. Thus, we ensured that response (if) received is from the middlebox and not the actual server. In the payload, we fudged the domain name and its offset within the request to determine exactly what triggered censorship. For *e.g.*, we set the HTTP Host field to that of an uncensored site, while the domain name of the censored site was positioned at a random offset within the HTTP header (say beyond the requested page indicated in the GET field). In all our tests we observed only when the Host field is set to the domain name of the censored site. Further details of more related experiments are presented in section 4.2.

---

[5] we used python difflib library for this purpose.

As described ahead in Section 4, in three of the four ISP where we observed HTTP censorship, the middlebox responds back to the client with a variant of the aforementioned censorship notification-cum-disconnection messages. These were mostly HTTP 200 OK responses carrying the state's censorship notification along with appropriate TCP bits enabled that force the client to terminate the connection with the PBW. They bore appropriate sequence and acknowledgement numbers (along with other protocol header information) to make them indistinguishable from legitimate packets which the client's underlying protocol stack expects, *wrt* the initial TCP connection to the PBW.

In Section 5, we show how we exploit this knowledge of protocol header idiosyncrasies, along with deliberate fudging of the requested domain name in the `Host` field of the `GET` requests to sidestep censorship.

**V. Identifying location of HTTP middleboxes:**. After characterizing the blocking behavior, we intend to identify the *network location* of middleboxes *viz.,* IP address. As earlier 3.2, we first ran `traceroute` to determine the number of hops between the client and a PBW (to be tested). We then use Iterative Network Tracing (shown in figure 1) whereby following a regular 3-way handshake to the PBW, we sent series a crafted HTTP `GET` request to it, with increasing TTL values until it encounters the middlebox whence the client observes a censorship-notification-cum-disconnection message (bearing TCP header with bits such as FIN or RST enabled). Correlating this TTL value against the IP address hops reported by `traceroute` helped us identify the middleboxes.

## 4. EXPERIMENTAL RESULTS

In order to determine the censorship mechanism we inspected for DNS, TCP/IP and HTTP blocking for the list of potentially blocked websites in nine major ISPs of the country (as explained in Section 3). For all ISPs, we found instances of DNS and HTTP filtering only.

### 4.1 DNS filtering

We began our study by identifying the open DNS resolvers in a chosen ISP. Thereafter applying our heuristics presented in Subsection 3.2 we determined which of the 1200 curated PBWs were being censored along with their corresponding DNS resolvers.

We observed poisoned DNS resolvers in only two of the nine ISPs, *viz.,* MTNL and BSNL.

Before presenting the results, we propose two metrics to analyze the extent of DNS filtering within the ISPs:

1. *Coverage*: Ideally all DNS resolvers of an ISP must be poisoned. We define the *coverage* as the fraction of all the resolvers of the ISP which are poisoned.

2. *Consistency*: Ideally the same set of sites must be blocked by all the *poisoned* resolvers of an ISP. We determined the set of filtered URLs as well as all the resolvers that blocked them. For every filtered URL we determine the fraction of poisoned resolver blocking it. *Consistency* is the average of these fractions.

In MTNL, we found a total of 448 resolvers, out of which 383 were poisoned, *i.e.* coverage was around 77%. Whereas, in BSNL we found only 17 poisoned resolvers out of a total of 182 (a smaller coverage of around 9.3%).

The consistency of each ISP can be inferred from Figure 2. Websites which are blocked in any of the two ISPs are represented on the X-axis. The percentage of resolvers blocking the website are represented on Y-axis. For the sake of preserving anonymity, we represent the sites with unique numbers, rather than actual names. It can be clearly seen from the figure that in general a single website (for *eg.,* website ID 450) is blocked by more number of resolvers in MTNL (44%) than in BSNL (6.6%). The consistency metric in MTNL (42.4%) is also higher than that of BSNL (7.5%).

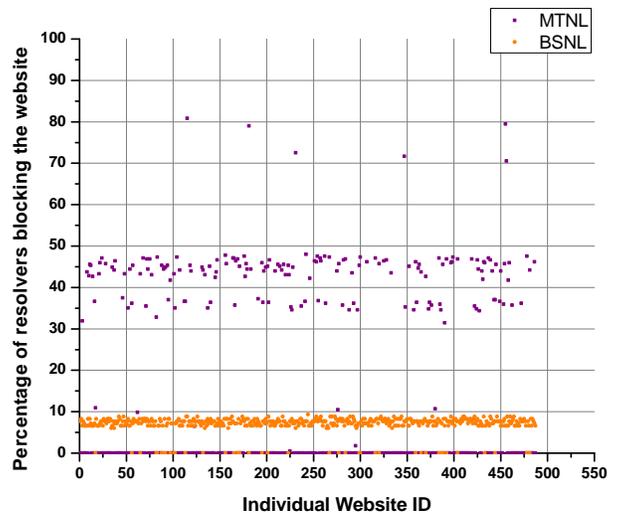

Figure 2: Consistency of DNS resolvers.

### 4.2 HTTP filtering

We found HTTP filtering in four out of nine ISPs. As already discussed in section 3.4, ISPs have deployed middleboxes which inspect the packets between the client and blocked websites with an intent to do blocking.

We began by identifying all those websites, among the 1200 PBWs, that was censored by the ISP. For instance in Airtel, we observed a total of 234 websites to be censored. The corresponding number for the remaining three ISPs is presented in the last column of table 2.

Using the approach described in subsection 3.4 we determined that censorship triggered *solely* due to request and not the response[6].

Finally, we attempt to find the actual network location of the middleboxes with a variant of our Iterative Network Tracing, involving crafted HTTP `GET` requests. However, we were unable to pinpoint the exact IP address of the middleboxes in most of our measurements because of anonymization by the ISP. We discuss this in detail in Section 6.

We now present in the behavior of the different types of middleboxes, we identified in the wild and describe their censorship mechanisms in detail.

### 4.2.1 Types of middleboxes

In our experiments, we identified two kinds of middleboxes–*viz.* *Interceptive Middlebox (IM)* and *Wiretap Middlebox (WM)*. IMs, hitherto unreported by others [49] are akin to transparent proxies which intercept connections between the client and server and establish their own to the server. In our studies, we found IMs which intercept client to PBW connections and send back censorship notification messages back to the client, without relaying the requests to the PBWs.

The other, *i.e.* WMs, have also observed in various other censorious regimes [49], and involve a host that is connected to an active network element via a wiretap. It receives a copy of all the packets exchanged and inspects for requests that need to be censored. Thereafter, it crafts responses and sends it back to the censor, with appropriate TCP header bits, with the intent to terminate existing connections.

The WMs are not as efficient as IMs, as they incur a higher cost in searching all flows to find the ones censor. They cannot outpace the client–PBW traffic flow, as they work with a copy of the packets, and therefore not as effective in filtering every single request with real-time efficiency (For WMs, in roughly 3 out of 10 attempts website gets rendered at the client machine, whereas for IMs all attempts to open the website were unsuccessful).

***Interceptive middleboxes.*** We used our variant of iterative network tracer (explained in Subsection 3.4) to first obtain the location of middlebox in the network path intervening the client and a filtered site (identified beforehand using `traceroute`).

Thereafter, we sent crafted HTTP `GET` requests with TTL values large enough to get past the network hop, corresponding to the middlebox. Regardless of any further increments to this IP TTL value, we never observed the expected ICMP TTL Expired responses, but rather received the censorship notification messages. In order to verify that triggering event occurs only for the blocked domain in `Host` field of `GET` request, we sent a crafted request where `Host` field's value was a non-censored domain, with iteratively increasing values. Interestingly, in these cases, we always received ICMP TTL Expired messages, even after TTL was large enough for the packets to transit the middlebox.

We went a step ahead and selected an array of hosts we controlled in different networks[7] outside Indian ISPs. On these machines, we hosted an ordinary webserver. From our client, hosted in the ISP under test, we created TCP connections to these remote machines. The remote host simultaneously also monitors its own traffic. The client sends crafted `GET` requests with `Host` field requesting a censored domain. The destination IP address, however, was that of the remote host. Upon traversing a censorious middlebox positioned on the intervening network path, the client receives a censorship notification packet-cum-disconnection packets, with TCP FIN bits enabled. The subsequent the 4-way disconnection always timeout (very likely dropped by the middlebox). Eventually, the client attempts to terminate the connection by sending a RST packet. The remote host receives none of these packets, other than the initial handshake messages and the said RST packet. The TCP sequence number of this RST packet differs from the one sent by the client, thereby confirming that it was sent by the middlebox.

Further, all packets with the source IP of client and destination IP of the blocked site are filtered after the initial `GET` request and before the final RST packet.

We repeated the same exercise, by replacing the `Host` field with that of an uncensored domain. Interestingly enough, the request reaches the remote host unfiltered.

The functioning of the interceptive middlebox can be is schematically shown in Figure 3.

***Wiretapping middleboxes.*** Similar to interceptive middleboxes, we used our own variant of iterative network tracer, to first obtain the location of the middlebox in the network path intervening the client and a filtered site.

After establishing TCP connection to the filtered site, the client sent HTTP `GET` requests to it that bore the URL of the blocked domain in the `Host` field. Inspecting the network traffic for the said message exchanges through `pcap`, we see that the client receives the censorship notification-cum-disconnection packet, with the forged IP address (that of the server) and TCP FIN+PSH bits enabled, which thereby enforces connection termination. But even before the termination process resolves, the client receives a fresh TCP RST packet from the middlebox, bearing the forged IP address of the server that forces the client to terminate the connec-tion immediately regardless of whether the termination process, which is underway, completes or not.

---

[6]Middleboxes gets triggered *solely* in the presence of filtered domains in Host field of the `GET` requests

[7]Planetlab, cloud services and hosts in different universities

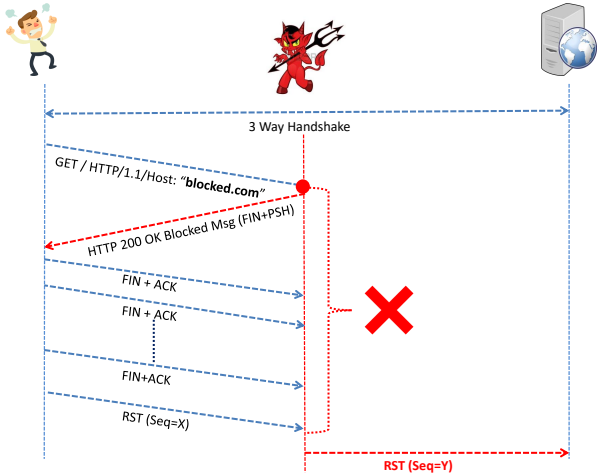

Figure 3: Censorship mechanism of interceptive middlebox.

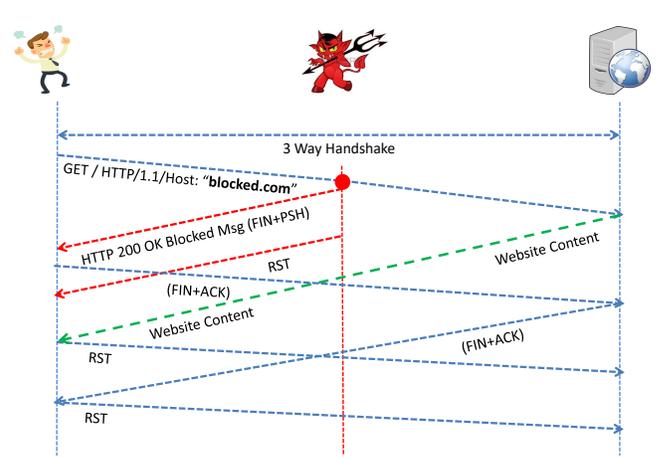

Figure 4: Censorship mechanism of Wiretapping middlebox.

Surprisingly, actual response from the filtered site eventually also reaches the client, but the connection to the server is already terminated by then. The client responds to packets with (as expected) with a TCP RST packet.

In order to confirm the censorship mechanism of wiretap middlebox, we adopted an approach similar to the one described for interceptive middleboxes, involving remote servers under our control. We sent crafted HTTP GET requests bearing a filtered domain, to our controlled remote server. These packets elicit the censorship notification-cum-disconnection messages, bearing the (forged) IP address of the remote host. The remote host, under our control, however, receives the GET request, but does not process them as it does not host the requested domain.

The behavior of the wiretap middleboxes is shown in Figure 4.

**Caveat:.** Are middleboxes stateful or do they inspect all packets? Our initial traffic inspections using pcap hint towards stateful middleboxes that commence traffic inspection only after complete TCP 3-way handshake is resolved.

To confirm our hunches we began with the client using traceroute command to record the number of network hops between itself and the filtered site. Thereafter the client sends a TCP SYN packet with TTL just large enough to get the packet to the penultimate hop (and *not* the destination), thus avoiding a full-fledged TCP 3-way handshake.

Finally, the client sends a crafted GET request whose Host field points to a filtered domain and that has the same TTL value, just used, so that it expires upon reaching the penultimate hop.

If the middleboxes commence traffic inspection upon observing every fresh TCP SYN packet, they must also then inspect the subsequent crafted GET request and respond back to the client with the censorship notification-cum-disconnection message. However, we never observed anything like that.

All other similar heuristics, such as starting by sending a SYN+ACK or not sending the final ACK of a regular 3-way handshake, but then sending the subsequent crafted GET request bore no different outcomes.

Finally, crafted HTTP GET request, bearing censorious domain requests in the Host field, but with no preceding TCP handshake, also does not seem to trigger censorship.

This confirms that the middleboxes are stateful and commences traffic inspection only when they observe a complete TCP handshake. These seem different from what were observed by Wang *et al.* [49] who looked into the architecture Chinese censorship infrastructure.

#### 4.2.2 Analyzing Extent of HTTP Filtering

In order to analyze the extent of HTTP filtering in an ISP we proposed variants of the two previous metrics, *viz. consistency* and *coverage*.

1. *Coverage*: A censorious ISP which is willing to use HTTP filtering ideally must deploy middleboxes in a manner, such that all the router-level paths inside an ISP must be intercepted by the middleboxes. We intend to find intra ISP router level paths which are intercepted by the middleboxes. *Coverage* is the fraction of all router-level paths scanned that are intercepted by middleboxes (we call *poisoned paths*).

2. *Consistency*: Ideally same number of websites must be blocked on all the poisoned paths of the ISP. In such

a case we say the ISP is 100% consistent. For every filtered URL we determine the fraction of poisoned paths blocking it. *Consistency* is the average of these fractions.

In order to find consistency and coverage we started our experiments with single vantage point (VP) in the ISPs. As already discussed earlier, HTTP censorship middleboxes are agnostic to the destination IP addresses of the HTTP `GET` requests (as long as they appear to be a part of an existing TCP connection). We harness this behavior of middlebox to find their coverage and placement statistics.

We establish TCP connections with Alexa top 1000 websites from the client machine and sent `GET` requests with `Host` fields pointing to all 1200 PBWs. Even if for single `GET` request we observed the censorship, we considered that path to be *poisoned* by the middlebox. For Reliance Jio ISP, we only observed 64 out of 1000 paths to be tainted with middlebox. This gave us the hint that maybe middleboxes are not placed optimally to intercept a large fraction of ISP paths.

Thus, to further test our observation with more vantage points, we used various hosts outside India, but under our control (PlanetLab nodes, cloud infrastructure, and some others in various universities). Our aim was to find the maximal possible number of middleboxes and the fraction of paths they intercept, inside an ISP.

For doing so, we began by scanning all live IP prefixes[8] for a particular ISP, and searched for hosts with open TCP port 80. Then we randomly sample two such IPs per prefix. We recorded the router-level path leading and the number of hops to each of these prefixes, from each vantage point, using `traceroute`.

We tailored our Iterative Network Tracing, targeting traces to each of these IPs (for all ISPs), where for each targeted host, we send 1200 HTTP `GET` requests, corresponding to each of the PBWs. If we get the censorship notification-cum-disconnection response for even a single site, we consider the corresponding network path to be *poisoned*.

We summarize our results in table 2. Column two and three represents coverage for an ISP from a single VP within ISP and multiple VPs outside of the ISP. Column four describes which type of middlebox (interceptive or wiretap) is deployed in the ISP and last column describes the total number of websites blocked out of 1200 PBW. It can be observed that Idea has highest coverage (90%) whereas Vodafone has very low coverage value (2.5%).

For Reliance Jio, we observed a very different behavior. While we observed a relatively low coverage of about 6.4% when searching for middleboxes from a vantage point positioned inside the network, we found no middleboxes when probing for middleboxes from remote hosts to hosts inside the ISP with open TCP port 80. We never observed a single instance of middlebox while we were probing from our distributed VPs. There are two possible explanations for this – either they have placed middleboxes sub-optimally and thus they are not observed to the VPs outside of ISP. Alternatively, they maybe filtering request not only on domain names but also for source IPs belonging to Jio network itself.

---
[8]obtained from CIDR report [2]

| ISP | Coverage (%) (**VP**: within ISP) | Coverage (%) (**VPs**: outside ISP) | Middle-Box Type | No. of websites blocked |
|---|---|---|---|---|
| Airtel | 75.2 | 54.2 | WM | 234 |
| Idea | 92 | 90 | IM | 338 |
| Vodafone | 11 | 2.5 | IM | 483 |
| Jio | 6.4 | 0 | WM | 200 |

**Table 2: HTTP filtering in different ISPs.**

Since, we never observed the IP address of the middlebox, we were unable to find the exact reason that why we never observed any middlebox when tested from outside of the ISP network.

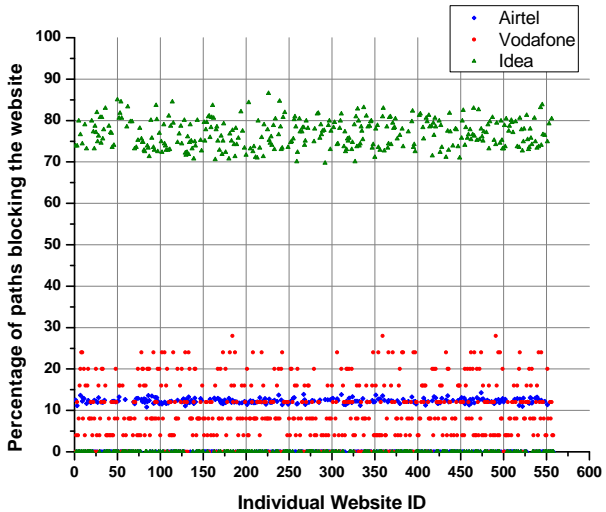

**Figure 5: Consistency of middleboxes.**

After finding the coverage of different ISPs, we now present the results obtained from computing consistency for each of them. In figure 5, X-axis represents websites which are blocked in any of the three ISPs (Vodafone, Airtel and Idea). The percentage of ISP paths that block a particular website are represented on Y-axis. It is evident from the figure, that on an average Idea network has highest consistency (76.8%) followed by Airtel (12.3%) and Vodafone (11.6%). It can be interpreted as, in Idea network a single website is blocked on 76.8% of the poisoned paths, as opposed to Airtel and Vodafone in which it is blocked by only ≈ 11 − 12% of the poisoned paths.

So far we discussed all details regarding HTTP filtering, but ignore HTTPS. We observed fewer than five instances of HTTPS filtering which were actually due to manipulated DNS responses by poisoned resolvers.

## 4.3 Collateral Damage within Indian ISPs

ISPs have contractual commercial agreements for routing Internet traffic among themselves [32] making them *neighbors*. Collateral damage occurs when traffic of non-censorious ISP transit through its peering censorious ISP. Previous studies that have explored censorship by collateral damage [39, 21], have focused on how one nation is impacted by the Internet censorship policies of other nation. Whereas, in this work we highlight collateral damage is possible within the same country itself. For instance, in ISPs like NKN, Sify and Siti, we never observed any filtering caused by their own policies, rather all the censorship instances are solely due to its peers policies, whereas for MTNL and BSNL it is the cumulative effect of its own and neighbors' policies. Table 3 summarizes our findings.

| ISPs (cesnored) | Neighboring ISPs (causing censorship) |
|---|---|
| NKN | Vodafone (69), TATA[9] (8) |
| Sify | TATA (142), Airtel (2) |
| Siti | Airtel (110) |
| MTNL | Airtel (25), TATA (134) |
| BSNL | Airtel (1), TATA (156) |

Table 3: **Collateral Damage: Non censorious ISP observe censorship due to their censorious peering ISPs. In NKN, we observed** 69 **websites were blocked by Vodafone and** 8 **were blocked by TATA communication.**

## 5. ANTI-CENSORSHIP APPROACHES

Broadly classifying we observed two types of censorships in popular ISPs of India *viz.,* HTTP filtering and DNS poisoning. In order to anti-censor them, we opted techniques depending upon the middlebox under action. Our solutions are simple and extremely effective.

**Evading *DNS poisoning:*** In order to circumvent poisoned DNS resolver, any non-poisoned resolvers can be used. We tested with OpenDNS, Google's public DNS (8.8.8.8) and many other non-poisoned resolvers which belong to non - censorious countries like Ireland, Canada, and Sweden. With each of them, we were able to bypass the DNS based censorship.

**Evading *HTTP filtering:*** As already explained in section 3.4 middlebox gets triggered upon identifying a blocked domain in the HOST field of GET request only. *Our goal is to craft such a GET request, which is not correctly interpreted by the middlebox but by the actual website.* We tried various techniques involving string fudging [36], such as manipulating the Host field values, prepending www to the website name, changing cases of the keywords like HTTP, GET and HOST, adding spaces before and after the domain name *etc..* Additionally we also tried approaches, like sending fragmented GET requests and using HTTP 2.0 as the underlying web protocol (instead of HTTP 1.1). Different approaches worked for sidestepping different middleboxes.

**I. Wiretapping middleboxes:** There are two approaches with which we bypassed these middleboxes.

- Changing the case of Host keyword in the GET request: Most popular browsers, like Mozilla Firefox and Google Chrome, use the title case for the Host keyword. Merely changing the case (*e.g.* changing it to HOst, HoST, HoSt or HOST*etc.*) was sufficient for request to go undetected via the middleboxes (of Airtel and Jio), but resulting in response from the actual blocked webserver. This suggests that the webservers, corresponding to the PBWs, adhere to RFC 2616 [8] and accept the keyword Host agnostic of the case, while the middleboxes look for exact keyword matches.

- Dropping the packets with RST or FIN bit set: As mentioned earlier, the censorship notification-cum-disconnection packet has the TCP FIN bit set. Subsequently the middlebox also sends a TCP RST packet to enforce the client to disconnect.

  Using iptables utility, all the packets (of blocked website's IP) which have FIN or RST bit set were dropped by the kernel. For Airtel, we observed that responses from middleboxes of Airtel always bear a fixed IP-Identifier value of 242. Thus, we added a general rule that FIN or RST packets with IP-Identifier field 242 must be dropped. This effectively filters the responses from the middleboxes.

  Since the actual GET requests are not dropped by the middlebox, they reached the blocked website and elicit regular responses. These response containing the actual content of the website and are accepted by the client browser.

**II. Interceptive middleboxes:** We further found two types of interceptive middleboxes *i.e.,* one which sends only censorship notification-cum-disconnection message to the client (overt) and other which sends only a RST packet to the client without any censorship notification (covert).

- *Overt Censorship*: To bypass such middleboxes which overtly censors the content, we fudged the Host field of the GET request. The standard domain request looks like "Host: blocked.com", *i.e.*

only one space between ':' and 'blocked.com'. But, instead, if we place additional spaces (or tab) in-between, *i.e.* "`Host:       blocked.com`", then the requests go undetected by the middleboxes, but servers interpret them correctly. Also, adding extra spaces (or tables) after the domain name works, *e.g.*"`Host:blocked.com        `".

- *Covert Censorship*: For bypassing such middlebox we intentionally inserted multiple `Host` fields (with different website names) in the same `GET` request to check which one of those is inspected by the middlebox. For all cases, we observed that it is triggered upon inspecting *only* the last `Host` keyword. Thus appending an uncensored domain request to the array of such `Host` keywords, we were able to bypass the middleboxes, but the server also neglects it as the request is not a standard one. Thus we crafted an unusual `GET` request, which looked something like "GET / HTTP/1.1 `Host:` blocked.com...\r\n\r\n `Host:` allowed.com". This request is neglected by the middlebox but on the other hand, accepted by the actual blocked website. Since middlebox is only looking at last `Host` keyword, it interprets that packet as non-suspicious and allows it to pass through. The server, on the other hand, treats the '\r\n\r\n' as the end of the GET request and the subsequent "`Host:` allowed.com" as a separate request. Thus, the client receives two responses from the website — the actual content which due to the first `Host` field and the BAD REQUEST message for the subsequent one.

## 6. DISCUSSION

### 6.1 Count of middleboxes in the ISP

In the previous study on China [52] authors reported that they found 495 router interfaces that have filtering device attached to them. However, in India, we could not follow the same approach. Throughout our research, we used `traceroute`s and iterative network tracer, with an intent of finding the *location* of censorship infrastructure. In all our tested ISPs, generally middlebox (or routers to which they are attached) show up as unresponsive routers (asterisked) when probed using `traceroute`. It is natural to ask if IP address of the middlebox is not known then how we confirmed that blocking behavior was an outcome of our test ISP or one its upstream provider? We applied few heuristics:

(1) On the paths where we observed the IP of a middlebox, first we confirmed that it belongs to same test ISP and later we recorded the corresponding censorship notification. If from other anonymized middleboxes we got the same notification, we considered it to be from same ISP.

(2) In path segments where asterisked router appeared between visible ones, we checked if the latter belonged to the same test ISP. If so then we assume that anonymized IPs belong to the same ISP.

(3) The censorship notification messages have unique characteristics *for eg.,* in Airtel, the censorship notification packet has an embedded `iframe` which redirects to *"airtel.com/dot"* and in Reliance JIO censored response redirects to its own unique IP address. In such cases, we easily identify the ISP of anonymized middlebox.

### 6.2 Issues with OONI

As already explained in section 3.1, OONI performs two sets of experiments for a given list of PBWs (1) accessing sites from client machine and (2) and accessing the same from a control sever (of OONI). If discrepancies in IP address resolutions (DNS censorship) or retrieved site contents (HTTP censorship) are observed, OONI flags the PBW to censored.

However, we found that results of OONI are *misleading*. It suffers from both false positives and false negatives. We now outline few possible reasons for false positives (incorrect flagging of sites as being censored):

- An Unavailable website, previously hosted on hosting services like GoDaddy, if removed, may give different HTTP responses when accessed from different locations – an artifact of distributed hosting. Though not a case of censorship, OONI flags them as filtered.

- Many websites have dynamic content such as live news feeds and advertisement embedded in the HTTP 200 OK messages that are often location dependent. These are also classified by OONI as being censored.

Also, OONI tool inspects differences in HTTP headers and body lengths of the response. If differences are greater than a threshold, it considers the site to be filtered. We observe that for a website hosted on CDN, the response at different geographic locations may come through different servers having obvious differences in the response metadata. In reality, such sites may not be blocked.

Thus when we created our scripts, we only calculated the difference in the content of the response, not the headers. If the difference is greater than the threshold, rather than directly reporting them as blocked, we manually verified them for blocking.

We now discuss why OONI often fails to detect a censored site (false negatives). In order to identify censorship, OONI calculates the difference between (1) the body length of compared websites (over the control server and the network of the user) differs by some percentage (2) the HTTP header names do not match and (3) the HTML title tags do not match.

*Even if a single aforementioned condition does not hold true [7, 12], OONI considers the website to be non censorious.* The following are few possible cases where OONI reports false negatives:

- We observed that for some websites, the response does not contain the content, rather a legitimate redirection link sent by the actual server. Similarly, in the censorship notification-cum-disconnection packets, there is an embedded `iframe` (which redirects to blocked page). For both the cases, the difference in the body length (of the responses) may be very less [10]. Thus, violating the condition one.

- OONI flags a website as non censored if the header fields (not their values) of both the HTTP response matches exactly[11]. In our measurements, we have observed that most of the middleboxes use the same HTTP header as of the actual web server. Thus, the headers of censorship notification-cum-disconnection packets matches with response header from the actual server. So, OONI mistakenly classifies a censored website to be non censored as the second condition is violated.

Censorship notification-cum-disconnection packets that we observed have no HTML title tags. On the contrary, most of the responses from the actual website did. OONI compares the title tags from both responses *only if* at least one word in both the tags is at least five characters long. Thus, in our case, this condition is never checked. Thus, only the two aforementioned conditions may contribute towards the false negatives.

### 6.3 Idiosyncrasy of middleboxes

- Ideal middleboxes inspect traffic agnostic of their port number, while all the rest inspect only requests destined to TCP port 80.

- WM specific to Airtel have a unique characteristic – all packets generated from these middleboxes have a fixed IP-ID value (242) in the IP header; whereas for all others', this is variable.

- There are some websites which are now unavailable (tested via Tor circuits ending in non-censorious country) but still blocked by the ISPs (both through HTTP and DNS filtering). This implies that ISPs are not updating their blacklists.

- Middlebox (IM and WM) maintains a state for all transiting TCP connections. It inspects all the connections for $2-3$ minutes, waiting for sensitive content[12] to arrive. If it does not receive any packet in that duration, it times out and purges the corresponding TCP state data. However, if fresh packets (corresponding to the individual flows, regardless of whether they are `GET` requests or not) arrive in the meantime, it restarts the timer for the timeout.

## 7. CONCLUDING REMARKS

In this work, we report a comprehensive analysis of censorship mechanism and infrastructure in nine popular ISPs of India. We commenced our research by using popular censorship detection tool OONI, but since we observed high false positives and negatives, we discontinued it. We developed our own automated approach (Interative Network Tracer), along with various heuristics, which we used to determine the type of censorship mechanism involved (and in some cases the approximate location of the censorship infrastructure as well). At every step we confirm our findings against the ground truth, an effort largely ignored by several others in the recent and distant past.

We primarily found DNS and HTTP filtering as the *only* techniques of censorship employed by these ISPs. Further, we evolved metrics, *viz. coverage* and *consistency* that respectively describe how well the censorship infrastructure covers the ISP and how consistent they are in censoring filtered domains. In passing, we also observed interesting cases of collateral damage within the ISPs of the same country. Finally, we developed novel anti-censorship techniques, involving local firewalling and manipulating the HTTP `GET` requests, through which we were able to bypass all forms of censorship without relying on conventional methods involving proxies and VPNs.

---

[10] Other variants of such scenarios are also possible *for eg.,* a small sign up/login page upon accessing the website
[11] We inspected the source code to ensure that this is the only case
[12] Domain names which are possibly blocked.